\documentclass
[superscriptaddress,secnumarabic,amssymb,amsmath,nobibnotes,aps,prd,showkeys,showpacs,nofootinbib,onecolumn,notitlepage,12pt]{revtex4}%
\usepackage{setspace}
\usepackage{color}
\usepackage{amsmath}
\usepackage{amsfonts}
\usepackage{verbatim}
\usepackage{amssymb}
\usepackage{graphicx,bm}
\usepackage{graphicx}
\usepackage{bbm}
\usepackage{amsmath}
\usepackage{amssymb}
\usepackage{amssymb}
\usepackage{graphicx,bm}
\usepackage{graphicx}
\usepackage{bbm}
\usepackage[caption=false]{subfig}%
\RequirePackage{latexsym}
\RequirePackage[colorlinks,citecolor=blue,urlcolor=blue,linkcolor=blue]{hyperref}
\providecommand{\U}[1]{\protect\rule{.1in}{.1in}}

\newcommand{\be}{\begin{equation}}
\newcommand{\ee}{\end{equation}}

\newcommand{\mincir}{\raise
-3.truept\hbox{\rlap{\hbox{$\sim$}}\raise4.truept\hbox{$<$}\ }}
\newcommand{\magcir}{\raise
-3.truept\hbox{\rlap{\hbox{$\sim$}}\raise4.truept\hbox{$>$}\ }}

\ifx\pdfoutput\relax\let\pdfoutput=\undefined\fi
\newcount\msipdfoutput
\ifx\pdfoutput\undefined\else
\ifcase\pdfoutput\else
\ifx\paperwidth\undefined\else
\ifdim\paperheight=0pt\relax\else\pdfpageheight\paperheight\fi
\ifdim\paperwidth=0pt\relax\else\pdfpagewidth\paperwidth\fi
\fi\fi\fi
\begin{document}
\title{Einstein-{\ae}ther Scalar-tensor Cosmology}
\author{Andronikos Paliathanasis}
\email{anpaliat@phys.uoa.gr}
\affiliation{Institute of Systems Science, Durban University of Technology, Durban 4000,
South Africa}
\affiliation{Instituto de Ciencias F\'{\i}sicas y Matem\'{a}ticas, Universidad Austral de
Chile, Valdivia 5090000, Chile}
\author{Genly Leon}
\email{genly.leon@ucn.cl}
\affiliation{Departamento de Matem\'{a}ticas, Universidad Cat\'{o}lica del Norte, Avda.
Angamos 0610, Casilla 1280 Antofagasta, Chile.}

\begin{abstract}
We propose an Einstein-{\ae}ther scalar-tensor cosmological model. In particular
in the scalar-tensor Action Integral we introduce the {\ae}ther field with {\ae}ther
coefficients to be functions of the scalar field. This cosmological model
extends previous studies on Lorentz-violating theories. For a spatially flat
Friedmann--Lema\^{\i}tre--Robertson--Walker background space we write the
field equations which are of second-order with dynamical variables the scale
factor and the scalar field. The physical evolution of the field equations
depends upon three unknown functions which are related to the scalar-tensor
coupling function, the scalar field potential and the {\ae}ther coefficient
functions. We investigate the existence of analytic solutions for the field
equations and the integrability properties according to the existence of
linear in the momentum conservation laws. We define a new set of variables in
which the dynamical evolution depends only upon the scalar field potential.
Furthermore, the asymptotic behaviour and the cosmological history is
investigated where we find that the theory provides inflationary eras similar
with that of scalar-tensor theory but with Lorentz-violating terms provided by
the {\ae}ther field. Finally, in the new variables we found that the field equations are integrable due to the existence of nonlocal conservation laws for arbitrary functional forms
of the three free functions.

\end{abstract}
\maketitle
\date{\today}

\section{Introduction}

\label{sec1}

A common property of various subjects of quantum gravity is the Lorentz
violation \cite{dav1}. Lorentz violation may not have been observed yet.
However, gravitational models which violate Lorentz symmetry have been of
special interest in the literature these last years
\cite{lorv1,lorv2,lorv3,lorv4,carroll,ea1,alm1,lv1,lv2,lv3}. A Lorentz
violating gravitational theory which has been widely studied is the
Einstein-{\ae}ther theory \cite{DJ,DJ2,Carru,esf}. In the Einstein-Hilbert Action
Integral there are introduced the kinematic quantities of a unitary time-like
vector field, the {\ae}ther field. The Lorentz symmetry is violated by the
definition of the preferred frame when the {\ae}ther field is selected. The limit
of Einstein's General Relativity exists while Einstein-{\ae}ther theory preserves
locality and covariance formulation. Einstein-{\ae}ther theory is a second-order
theory and it has been used for the description of various gravitational
systems \cite{ee1,ee2,ee3,ee4,ee5,ee6,ee7,ee8}. An important characteristic of
Einstein-{\ae}ther theory is that it can describes the classical limit of
Ho\v{r}ava-Lifshitz \cite{hor3}.  The inverse is not true, see the discussion
in \cite{st01a,st01b}.

On the other hand, scalar fields play an important role for the description of
the universe. The main mechanism for the description of the early acceleration
era of the universe is attributed to a scalar field, the inflaton.
Additionally, scalar fields have been used to describe the late-time
acceleration as possible solutions to the dark energy problem
\cite{guth,Ratra,Barrow,Linder,sf4,sf5,sf6,sf7}. Another important feature of the
scalar fields is that they can attribute the degrees of freedom provided by
higher-order derivatives in gravitational Action Integral as a result of
quantum corrections or modifications of General Relativity \cite{sf1,sf2,sf3}.
Consequently, gravitational models where they describe {\ae}ther Lorentz
violating inflationary solutions have been introduced in the literature. A
first attempt proposed by Kanno and Soda in \cite{Kanno:2006ty} while a more
general consideration was investigated by Donnelly and Jacobson in \cite{DJ}. A
common characteristic of these studies is that the scalar field has been
defined to be minimally coupled to gravity. In this work we extend the
previous considerations and specifically the model proposed in
\cite{Kanno:2006ty} for which we assume that the scalar field is nonminimally
coupled to gravity, that is, we select as natural frame the Jordan frame.\ We
shall call this model Einstein-{\ae}ther scalar-tensor theory.

The commonest scalar-tensor theory is the Brans-Dicke theory. It was
proposed in \cite{bd} and it is inspired by Mach's Principle. In Mach's
Principle, gravity is described by the metric tensor and by a scalar field
nonminimally coupled to gravity. The field plays an important role in the
description of the early universe \cite{ssen} and in general on the
construction of the physical space. For a debate on which is the natural
frame, the Jordan frame or the Einstein frame we refer the reader in
\cite{fj1,fj2,fj3} and references therein. Another well-known scalar-tensor
model is the dilaton theory, which is the fundamental Action Integral for
string cosmology \cite{s1}.

In this study we investigate the effects of the introduction of the aether
field in scalar-tensor theory in cosmology. For the description of the natural
space we assume an isotropic and homogeneous universe described by the
spatially flat Friedmann--Lema\^{\i}tre--Robertson--Walker \ (FLRW) metric.
For this spacetime, and for the comoving time-like aether field we derive the
field equations which are of second-order as in scalar-tensor theory. The
field equations depend on three unknown variables which are constructed by the
scalar field potential, the scalar field coupling function to gravity and the
coefficient functions which relates the aether and the scalar fields. We
investigate the existence of integrable cosmological models by defining new
variables and by using the Noether's symmetry approach \cite{ns1}. The latter
approach has applied for the classification of various cosmological models and
the determination of new analytic solutions \cite{ns2,ns3,ns4,ns5}.

Moreover, we study the cosmological asymptotic solutions provided by this
model by determining the stationary points of the field equations
\cite{ds1,ds2,ds3}. Such analysis is essential in order to infer about the
viability of the model \cite{ds4}. We found that the Einstein-aether
scalar-tensor model can provide more than one inflationary eras, as also it
provides a richer cosmological history in contrary to the scalar-tensor
theory. The plan of the paper is as follows.

In Section \ref{sec2} we define the model of our consideration and we derive
the cosmological field equations. In Section \ref{sec3} we define new
variables in which we are able to reduce the cosmological field equations into
a Newtonian integrable model. The asymptotic dynamics are investigated in
Section \ref{sec4}. Finally the symmetry classification scheme and the
derivation of a new analytic solution is presented in Section \ref{sec5}.
Finally, in Section \ref{sec6} we draw our conclusions.

\section{Einstein-aether scalar-tensor cosmology}

\label{sec2}

Einstein-aether cosmological models with scalar field have been introduced
before in the literature. In \cite{DJ} it has been considered the scalar field
potential for a quintessence field to be an arbitrary function of the
kinematic invariants of the aether field. This is a generic model which has
been used as a base model of study. The attention has been drawn by the
proposed\ Lagrangian by Kanno and Soda \cite{Kanno:2006ty}. They introduced a
gravitational Action integral in which the Einstein-aether coupling parameters
are functions of the scalar field. Such cosmological model admits two stages
for the inflationary era, the slow-roll era when the scalar field dominates,
and a Lorentz violating state when the aether field contributes in the
cosmological fluid.

The model proposed in \cite{Kanno:2006ty} affects the dynamics of the chaotic
inflationary model. It has been used as a toy model for the definition of a
Lorentz violating DGP model with no ghosts \cite{j1}. Lorentz violating
inflationary models has been widely studied in the literature, see for
instance \cite{j2,j3,j4,j5}. In \cite{j6}, the authors investigated the
effects of Lorentz violation in the cosmological history. Moreover, in
\cite{j4} the Lorentz violation in cosmology was constraint on the
cosmological observations. It was found that Einstein-\ae ther cosmology can
explain the cosmological observations.

The dynamics of cosmological models with an aether field was the subject of
study for many studies. In \cite{j7} the dynamics of the model proposed in
\cite{Kanno:2006ty} was investigated in details. For a purely scalar field it was
found that there are two attractor solutions which can describe inflationary
epochs, that is in agreement with the two inflationary eras as described in
\cite{Kanno:2006ty}. Other studies on the dynamics of Einstein-aether scalar
field models can be found in \cite{j8,j9,j10,j11,j12,j13,j10a}.

Exact and analytic solutions in Einstein-aether scalar field cosmology were
found by using the symmetry analysis in \cite{j14}. Moreover, the first
attempt to quantize in Einstein-aether scalar field cosmology was performed in
\cite{j15}. Specifically, by using the minisuperspace description was proposed
a canonical quantization approach where the Wheeler-DeWitt equation was defined
and was solved by investigating the existence of quantum operators which keep
invariant the Wheeler-DeWitt equation.

In this study we generalize the gravitational model proposed in
\cite{Kanno:2006ty} and assume that the scalar field is defined in the Jordan
frame, that is, the field $\phi\left(  x^{\nu}\right)  $ is coupled to
gravity. We propose the Einstein-aether scalar-tensor gravitational model
defined by the Action Integral
\begin{equation}
S=S_{ST}+S_{Aether}, \label{ac.01}%
\end{equation}
where $S_{ST}$ is the Action Integral for the scalar-tensor theory \cite{qq1}%

\begin{equation}
\int dx^{4}\sqrt{-g}\left(  F\left(  \phi\right)  R+\frac{1}{2}g^{\mu\nu}%
\phi_{;\mu}\phi_{;\nu}+V\left(  \phi\right)  \right)  ,
\end{equation}
function $F\left(  \phi\right)  $ is the coupling function of the scalar field
$\phi\left(  x^{\nu}\right)  $ with gravity which we assume that it is
nonconstant, and $V\left(  \phi\left(  x^{\nu}\right)  \right)  \,\ $is the
scalar field potential. The second term, $S_{Aether}~$in expression
\eqref{ac.01} corresponds to the aether field $u^{\mu}$ as defined in
\cite{Kanno:2006ty}
\begin{equation}%
\begin{split}
S_{Aether}=-\int dx^{4}\sqrt{-g}\Big[  &  \beta_{1}\left(  \phi\right)
u^{\nu;\mu}u_{\nu;\mu}+\beta_{2}\left(  \phi\right)  \left(  g^{\mu\nu}%
u_{\mu;\nu}\right)  ^{2}+\beta_{3}\left(  \phi\right)  u^{\nu;\mu}u_{\mu;\nu
}\\
&  +\beta_{4}\left(  \phi\right)  u^{\mu}u^{\nu}u_{;\mu}u_{\nu}-\lambda\left(
u^{\mu}u_{\nu}+1\right)  \Big].
\end{split}
\end{equation}

Functions $\beta_{1}\left(  \phi\right)  $,~$\beta_{2}\left(  \phi\right)
$,~$\beta_{3}\left(  \phi\right)  $ and $\beta_{4}\left(  \phi\right)  $ are
the coefficient functions which define the coupling between the aether field
and the scalar field. Moreover, $\lambda$ is a Lagrange multiplier which
ensure the unitarity of the aether field, i.e. $u^{\mu}u_{\mu}+1=0$.

According to the Cosmological principle, in large scale the universe is
assumed to be isotropic and homogeneous described by the spatially flat FLRW
line element
\begin{equation}
ds^{2}=-N^{2}\left(  t\right)  dt^{2}+a^{2}\left(  t\right)  \left(
dx^{2}+dy^{2}+dz^{2}\right)  , \label{ac.03}%
\end{equation}
in which $N\left(  t\right)  $ is the lapse function, $a\left(  t\right)  $ is
the scale factor and describes the radius of the three-dimensional Euclidean
space. For the comoving observer, the expansion rate $\theta$ is defined as
$\theta=3H^{2},$ where $H=\frac{\dot{a}}{Na}$ is the Hubble function and dot
means total derivative with respect to the independent variable $t$.

For the line element \eqref{ac.03} we calculate the Ricciscalar $R=12H^{2}%
+\frac{6}{N}\dot{H}$. For the scalar field and the Aether field we assume that
they inherits the symmetries of the background space, $\phi=\phi\left(
t\right)  $, while the unitarity condition for the Aether field provides
$u^{\mu}=\frac{1}{N}\delta_{t}^{\mu}$. We replace in \eqref{ac.01} and
integrating by parts the gravitational Action integral is written in the
minisuperspace description%
\begin{equation}
S=\int\frac{dt}{N}\left(  6F\left(  \phi\right)  a\dot{a}^{2}+6F_{,\phi}%
a^{2}\dot{a}\dot{\phi}+\frac{1}{2}a^{3}\dot{\phi}^{2}-N^{2}a^{3}V\left(
\phi\right)  +3\left(  \beta_{1}\left(  \phi\right)  +3\beta_{2}\left(
\phi\right)  +\beta_{3}\left(  \phi\right)  \right)  a\dot{a}^{2}\right)  .
\label{st.01}%
\end{equation}

Hence, the point-like Lagrangian which describes the field equations is%
\begin{equation}
L\left(  N,a,\dot{a},\phi,\dot{\phi}\right)  =\frac{1}{N}\left(  6A\left(
\phi\right)  a\dot{a}^{2}+6B\left(  \phi\right)  a^{2}\dot{a}\dot{\phi}%
+\frac{1}{2}a^{3}\dot{\phi}^{2}\right)  -Na^{3}V\left(  \phi\right)  ,
\label{st.02}%
\end{equation}
where the new functions $A\left(  \phi\right)  $ and $B\left(  \phi\right)  $
are defined as $A\left(  \phi\right)  =F\left(  \phi\right)  +\frac{1}%
{2}\left(  \beta_{1}\left(  \phi\right)  +3\beta_{2}\left(  \phi\right)
+\beta_{3}\left(  \phi\right)  \right)  $ and $B\left(  \phi\right)  =F\left(
\phi\right)  _{,\phi}$. The minisuperspace description it is an important
feature of gravitational models. The field equations describe the evolution of
point-like particles while methods from Analytic Mechanics can be applied.
Moreover, the existence of the minisuperspace is essential for the canonical
quantization of the theory which leads to the Wheeler-DeWitt equation of
quantum cosmology. In this work we are interested on classical solutions for
the field equations described by the singular point-like Lagrangian
\eqref{st.02}.

Variation with respect to dependent variables $N,~a$ and $\phi$ of the Action
Integral \eqref{st.01} provides the cosmological field equations
\begin{equation}
\frac{1}{N^{2}}\left(  6A\left(  \phi\right)  a\dot{a}^{2}+6B\left(
\phi\right)  a^{2}\dot{a}\dot{\phi}+\frac{1}{2}a^{3}\dot{\phi}^{2}\right)
+a^{3}V\left(  \phi\right)  =0, \label{st.03}%
\end{equation}%
\begin{equation}
2a\ddot{a}+\dot{a}^{2}-2a\dot{a}\frac{\dot{N}}{N}+2a\dot{a}\dot{\phi}%
-\frac{a^{2}\left(  \frac{1}{2}\dot{\phi}^{2}-N^{2}V\left(  \phi\right)
\right)  }{2A}+\frac{a^{2}B}{A}\left(  \frac{B_{,\phi}}{B}\dot{\phi}^{2}%
+\ddot{\phi}-\dot{\phi}\frac{\dot{N}}{N}\right)  =0, \label{st.04}%
\end{equation}%
\begin{equation}
\ddot{\phi}-\frac{\dot{N}}{N}\dot{\phi}+\frac{3}{a}\dot{a}\dot{\phi}+\frac
{6}{a^{2}}B\left(  \phi\right)  \left(  a\ddot{a}+2\dot{a}^{2}-a\dot{a}\dot
{N}\right)  -6A\left(  \phi\right)  _{,\phi}\left(  \frac{\dot{a}}{a}\right)
^{2}+N^{2}V_{,\phi}=0. \label{st.05}%
\end{equation}

Equivalently,%
\begin{equation}
6A\left(  \phi\right)  H^{2}+6B\left(  \phi\right)  H\frac{\dot{\phi}}%
{N}+\frac{\dot{\phi}^{2}}{2N^{2}}+V\left(  \phi\right)  =0 \label{st.06}%
\end{equation}%
\begin{equation}
\left(  \frac{2}{N}\dot{H}+3H^{2}\right)  +2\frac{A_{,\phi}}{A}H\frac
{\dot{\phi}}{N}-\frac{\left(  \frac{1}{2N^{2}}\dot{\phi}^{2}-V\left(
\phi\right)  \right)  }{2}+\frac{B}{A}\left(  \frac{B_{,\phi}}{B}\dot{\phi
}^{2}+\ddot{\phi}-\dot{\phi}\frac{\dot{N}}{N}\right)  =0 \label{st.07}%
\end{equation}%
\begin{equation}
\ddot{\phi}-\frac{\dot{N}}{N}\dot{\phi}+3H\dot{\phi}+6B\left(  \phi\right)
\left(  \dot{H}+3NH^{2}\right)  -6A\left(  \phi\right)  _{,\phi}H^{2}%
+N^{2}V_{,\phi}=0. \label{st.08}%
\end{equation}

Therefore, the field equations \eqref{st.06} and \eqref{st.07} can be written
as
\begin{align}
3H^{2}  &  =G_{eff}\rho_{eff},\label{st.09}\\
\frac{2}{N}\dot{H}+3H^{2}  &  =G_{eff}p_{eff}, \label{st.10}%
\end{align}
in which $\rho_{eff}$ and $p_{eff}$ are the energy density and pressure for
the effective fluid defined as%
\begin{align}
\rho_{eff}  &  =\frac{1}{2}\left(  6B\left(  \phi\right)  H\frac{\dot{\phi}%
}{N}+\frac{\dot{\phi}^{2}}{2N^{2}}+V\left(  \phi\right)  \right)
,\label{st.11}\\
p_{eff}  &  =-\left(  2A\left(  \phi\right)  _{,\phi}H\frac{\dot{\phi}}%
{N}-\frac{1}{2}\left(  \frac{1}{2N^{2}}\dot{\phi}^{2}-V\left(  \phi\right)
\right)  +B\left(  \phi\right)  _{,\phi}\frac{\dot{\phi}^{2}}{N^{2}}+B\left(
\phi\right)  \left(  \ddot{\phi}-\dot{\phi}\frac{\dot{N}}{N}\right)  \right)
, \label{st.12}%
\end{align}
and $G_{eff}=\frac{1}{\left(  -A\left(  \phi\right)  \right)  }$ is the
time-dependent gravitational constant. We remark that in contrary to the
scalar-tensor theory in which the effective gravitational constant depends on
the coupling function $F\left(  \phi\right)  $, in the Einstein-aether scalar
tensor model it depends also on the the aether coefficients functions
$\beta_{1}-\beta_{4}$. In addition we observe that if the aether coefficient
functions are constant, the usual scalar-tensor theory is recovered.

We continue our analysis by investigate the existence of analytic solutions
for the field equations \eqref{st.06}-\eqref{st.08}.

\section{Analytic solution}

\label{sec3}

Before we proceed,  we can define without loss of generality the new scalar
field $\psi$ with the relation $d\phi=\sqrt{A\left(  \psi\right)  }d\psi$.
Hence, the point-like Lagrangian \eqref{st.02} reads%
\begin{equation}
L\left(  N,a,\dot{a},\psi,\dot{\psi}\right)  =\frac{1}{N}\left(  6A\left(
\psi\right)  a\dot{a}^{2}+6B\left(  \psi\right)  a^{2}\dot{a}\dot{\psi}%
+\frac{A\left(  \psi\right)  }{2}a^{3}\dot{\psi}^{2}\right)  -Na^{3}V\left(
\psi\right)  . \label{st.14}%
\end{equation}

The field equations of this cosmological model depend on three arbitrary
functions, $A\left(  \psi\right)  ,~B\left(  \psi\right)  $ and $V\left(
\psi\right)  $ which should be defined. For the lapse function we
consider$~N=n\left(  t\right)  a^{-3}A\left(  \psi\right)  $, the point-like
Lagrangian \eqref{st.14} is written as follows%
\begin{equation}
L\left(  n,a,\dot{a},\psi,\dot{\psi}\right)  =\frac{a^{3}}{n\left(  t\right)
}\left(  6a\dot{a}^{2}+6\beta\left(  \psi\right)  a^{2}\dot{a}\dot{\psi}%
+\frac{1}{2}a^{3}\dot{\psi}^{2}\right)  -n\left(  t\right)  V\left(
\psi\right)  A\left(  \psi\right)  , \label{st.15}%
\end{equation}
in which $B\left(  \psi\right)  =\beta\left(  \psi\right)  A\left(
\psi\right)  $.

We continue by defining the new variable $X=\ln\left(  a\right)  +\frac{1}%
{2}\int\beta\left(  \psi\right)  d\psi$, that is $a=\exp\left(  X-\frac{1}%
{2}\int\beta\left(  \psi\right)  d\psi\right)  $, then Lagrangian function
\eqref{st.15} becomes%
\begin{equation}
L\left(  n,X,\dot{X},\psi,\dot{\psi}\right)  =\frac{1}{2n}e^{-6X-3\int
\beta\left(  \psi\right)  d\psi}\left(  12\dot{X}^{2}+\left(  1-3\beta\left(
\psi\right)  ^{2}\right)  \dot{\psi}^{2}\right)  -n\left(  t\right)  V\left(
\psi\right)  A\left(  \psi\right)  \label{st.16}%
\end{equation}

We select again the new lapse function $n\left(  t\right)  =e^{-6X-3\int
\beta\left(  \psi\right)  d\psi}$, in which the point-like Lagrangian reads%
\begin{equation}
L\left(  n,X,\dot{X},\psi,\dot{\psi}\right)  =\frac{1}{2n}e^{-6X}\left(
12\dot{X}^{2}+\left(  1-3\beta\left(  \psi\right)  ^{2}\right)  \dot{\psi}%
^{2}\right)  -e^{-6X}V\left(  \psi\right)  A\left(  \psi\right)
e^{-3\int\beta\left(  \psi\right)  d\psi}%
\end{equation}
Now for the scalar field potential $V\left(  \psi\right)  =V_{0}\left(
A\left(  \psi\right)  \right)  ^{-1}e^{3\int\beta\left(  \psi\right)  d\psi}$
the field equations are
\begin{align}
\left(  12\dot{X}^{2}+\left(  1-3\beta\left(  \psi\right)  ^{2}\right)
\dot{\psi}^{2}\right)  +2V_{0}e^{6X}  &  =0,\label{st.17}\\
\ddot{X}+V_{0}e^{6X}  &  =0,\label{st.18}\\
\ddot{\psi}-3\beta\left(  \psi\right)  \left(  \beta_{,\psi}\dot{\psi}%
^{2}+\beta\left(  \psi\right)  \ddot{\psi}\right)   &  =0. \label{st.19}%
\end{align}
while the element for the background space is%
\begin{equation}
ds^{2}=-\exp\left(  -12X\right)  A\left(  \psi\right)  dt^{2}+\exp\left(
2X-\int\beta\left(  \psi\right)  d\psi\right)  \left(  dx^{2}+dy^{2}%
+dz^{2}\right)  , \label{st.20}%
\end{equation}

The system of differential equations \eqref{st.17}-\eqref{st.19} describes the
motion of particle in the two-dimensional minisuperspace with line element%
\begin{equation}
ds_{\left(  ms\right)  }^{2}=\left(  12dX^{2}+\left(  1-3\beta\left(
\psi\right)  ^{2}\right)  d\psi^{2}\right)  , \label{st.21}%
\end{equation}
and effective potential $V_{eff}=V_{0}e^{6X}$.

The analytic solution of the field equations \eqref{st.17}, \eqref{st.18}, \eqref{st.19} is%
\begin{equation}
X\left(  t\right)  =\frac{1}{6}\ln\left(  \cosh\left(  \frac{\sqrt{3}t}{X_{0}%
}+X_{2}\right)  \right)  -\frac{1}{3}\ln\left(  X_{0}\right)  ,
\end{equation}%
\begin{equation}
\Psi\left(  t\right)  =\Psi_{1}t+\Psi_{0},
\end{equation}
with $d\Psi=\sqrt{1-3\beta\left(  \psi\right)  }d\psi$.

Consequently, for the integrability of the field equations the following
proposition follows.

\textit{Proposition 1: The cosmological field equations for the
Einstein-aether Scalar-tensor described by the point-like Lagrangian are
superintegrable for arbitrary functions }$A\left(  \psi\right)  $\textit{ and
}$B\left(  \psi\right)  $\textit{ when }$V\left(  \psi\right)  =V_{0}\left(
A\left(  \psi\right)  \right)  ^{-1}e^{3\int\frac{B\left(  \psi\right)
}{A\left(  \psi\right)  }d\psi}$\textit{. The field equations describe the
motion of a particle in the two-dimensional flat space in Cartesians
coordinates for the lapse function }$N\left(  t\right)  =\exp\left(
-12X\right)  A\left(  \psi\right)  $\textit{ with }$X=\ln\left(  a\right)
+\frac{1}{2}\int\frac{B\left(  \psi\right)  }{A\left(  \psi\right)  }d\psi.$

In the special case in which $\frac{B\left(  \psi\right)  }{A\left(
\psi\right)  }=const$, then for the scalar field potential and the coupling
function $A\left(  \psi\right)  $ follows $V\left(  \psi\right)  =\bar{V}%
_{0}A\left(  \psi\right)  ^{-1}$.

The determination of integrability for a given cosmological model, and in
general for a physical system is an important property. Nowadays, it is
popular to solve a dynamical system by using numerical techniques. However,
there are two important issues in that approach. When we solve a dynamical
system numerically we do not know if the evolution it is sensitive on the
initial conditions, especially when chaos exists. Moreover it is unknown if
the numerical solution corresponds to an actual solution on the problem. These
issues are solved when we determine the integrability of the given dynamical system.

In cosmology, the knowledge that a dynamical system is integrable is of
special interests. Integrable cosmological models have been found that they
can describe various areas in the evolution of the cosmological history.
Moreover, the initial value problem for inflation \cite{lin1} can be solved
easily for integrable models, while we know that the provided cosmological
history is not sensitive on the small changes of the initial conditions.
Although, an integrable model may not describe the complete cosmological
history, it can be used always as reference model. For instance, the
multi-body gravitational system is a chaotic dynamical system, the integrable
two-body system it describes well the Sun-Earth orbits.

In order to understand the general evolution of the given cosmological model
in the following section we investigate the asymptotic behaviour of the field equations.

\section{Asymptotic dynamics}

\label{sec4}

Let us perform a detailed study on the evolution of the field equations in the
Einstein-aether scalar-tensor model. We prefer to work with the scalar field
$\psi$, where the field equations are described by the Lagrangian function
\eqref{st.14}. We select to work in the $H$-normalization \cite{ds1} and to
define the new dimensionless variables%
\begin{equation}
x=\frac{\dot{\psi}}{2\sqrt{3}H}~,~y=\frac{V\left(  \psi\right)  }{A\left(
\psi\right)  H^{2}}~,~\lambda=\frac{V_{,\psi}\left(  \psi\right)  }{V\left(
\psi\right)  }\text{.} \label{st.23}%
\end{equation}
We remark that in general someone should consider a more general consideration
which will allow for the Hubble function to take the value zero. Here, we work
on the branch $H>0$ and we focus on the existence of asymptotic solutions
which describe de Sitter universes or scaling solutions.

For the arbitrary functions of the dynamical system we assume $B\left(
\psi\right)  =\frac{\beta_{0}}{\sqrt{3}}A\left(  \psi\right)  ,~\beta_{0}%
^{2}\neq1$, with $A\left(  \psi\right)  =A_{0}e^{-\nu\psi}$. Therefore, in the
new variables the field equations read%
\begin{equation}
y=-6\left(  1+2\beta_{0}x+x^{2}\right)  \label{st.24}%
\end{equation}
and%
\begin{equation}
\frac{dx}{d\ln a}=\frac{1}{1-\beta_{0}^{2}}\left(  \sqrt{3}\left(  \lambda
+\nu\right)  \left(  1+\beta_{0}x\right)  -3\left(  \beta_{0}+x\right)
\right)  \left(  1+2\beta_{0}x+x^{2}\right)  \label{st.25}%
\end{equation}%
\begin{equation}
\frac{d\lambda}{d\ln a}=2\sqrt{3}\lambda^{2}x\left(  \Gamma\left(
\lambda\right)  -1\right)  ~,~\Gamma\left(  \lambda\right)  =\frac
{V_{,\psi\psi}V}{\left(  V_{\psi}\right)  ^{2}}. \label{st.26}%
\end{equation}

Every stationary point for the dynamical system \eqref{st.25}-\eqref{st.26}
describes an exact solution for the scale factor for the background space.
Indeed, the effective equation of state $w_{eff}=\frac{p_{eff}}{\rho_{eff}}$
in the new variables is%
\begin{equation}
w_{eff}\left(  x,\lambda\right)  =-1+\frac{2\beta_{0}\left(  3\beta_{0}%
-\sqrt{3}\lambda\right)  -2\left(  \sqrt{3}\beta_{0}\lambda-3\right)  \left(
2\beta_{0}+x\right)  x-2\sqrt{3}\nu\left(  \beta_{0}+x\left(  2+\beta
_{0}x\right)  \right)  }{3\left(  1-\beta_{0}^{2}\right)  }. \label{st.27}%
\end{equation}
Hence, at the stationary point $P$ with coordinates $P=\left(  x\left(
P\right)  ,\lambda\left(  P\right)  \right)  $, $w_{eff}\left(  P\right)
=const$, that is, by definition $a\left(  t\right)  =a_{0}t^{\frac{2}{3\left(
1+w_{eff}\left(  P\right)  \right)  }}$ for $w_{eff}\left(  P\right)  \neq-1$
and $a\left(  t\right)  =a_{0}e^{H_{0}t}$ for $w_{eff}=-1$. Consequently, when
we determine the stationary points of the system \eqref{st.25}-\eqref{st.26}
and investigate their stability we are able to infer about the asymptotic
behaviour for the Einstein-aether scalar-tensor model.

For the scalar field potential we consider the following potential function
$V\left(  \psi\right)  =V_{0}e^{\lambda\psi}$.

\subsection{Exponential potential}

For the exponential potential, $V\left(  \psi\right)  =V_{0}e^{\lambda\psi}$,
parameter $\lambda$ is always a constant, thus the dynamical system is reduced
to the one dimension equation \eqref{st.25}. The right hand since of equation
\eqref{st.25} vanishes when
\[
x_{1}=-\frac{3\beta_{0}-\sqrt{3}\left(  \lambda+\nu\right)  }{3-\sqrt{3}%
\beta_{0}\left(  \lambda+\nu\right)  },\]
\[x_{2}=-\beta_{0}+\sqrt{\beta_{0}%
^{2}-1},\] \[x_{3}=-\beta_{0}-\sqrt{\beta_{0}^{2}-1}.%
\]
Consequently, we determine three stationary points, namely $A_{1},~A_{2}$ and
$A_{3}$ with coordinates $x_{1}$, $x_{2}$ and $x_{3}$ respectively.

For point $A_{1}$ we derive $y\left(  A_{1}\right)  =\frac{18\left(
1-\beta_{0}^{2}\right)  \left(  2\sqrt{3}\beta_{0}\left(  \lambda+\nu\right)
-3-\left(  \lambda+\nu\right)  ^{2}\right)  }{\left(  3-\sqrt{3}\beta\left(
\lambda+\nu\right)  \right)  ^{2}}$, and $w_{eff}\left(  A_{1}\right)
=-\frac{3+2\left(  \lambda^{2}-\nu^{2}\right)  +\sqrt{3}\beta_{0}\left(
\nu-3\lambda\right)  }{\sqrt{3}\beta_{0}\left(  \lambda+\nu\right)  -3}$.
Therefore, the asymptotic solution at $A_{1}$ describes a universe where the
kinetic part, the potential part of the scalar field as also the aether field
contribute in the cosmological fluid. The scalar factor is scaling, except
from the case in which $\nu=\sqrt{3}\beta_{0}-\lambda$ or $\nu=\lambda$ where
the asymptotic solution describes a de Sitter universe and $w_{eff}\left(
A_{1}\right)  =-1$. The linearized system around the stationary point $A_{1}$
becomes $\frac{d\delta x}{d\ln a}=e_{1}\left(  \lambda,\nu,\beta_{0}\right)
\delta x$ in which $e_{1}\left(  \lambda,\nu,\beta_{0}\right)  =\frac{3\left(
3+\left(  \lambda+\nu\right)  \left(  \lambda+\nu-2\sqrt{3}\beta_{0}\right)
\right)  }{\sqrt{3}\beta_{0}\left(  \lambda+\nu\right)  -3}$. When $e_{1}<0$
the stationary point is an attractor and the asymptotic solution is stable. 
Otherwise, the point $A_{1}$ is a source.\ Let us focus in the special case
that we studied before for the integrable model in which $\nu=-\lambda$. In
this case we calculate $y\left(  A_{1}\right)  =-6\left(  1-\beta_{0}%
^{2}\right)  ,~w_{eff}\left(  A_{1}\right)  =-1+\frac{4}{3}\beta_{0}\lambda$
and $e_{1}\left(  \lambda,-\lambda,\beta_{0}\right)  =-3$. We conclude that
the stationary point is always an attractor and $w_{eff}\left(  A_{1}\right)
<-\frac{1}{3}$ when $\beta_{0}\lambda<\frac{1}{2\sqrt{3}}$. In addition, $y>0$
when $\beta_{0}^{2}>1$. On the other hand, for the two de Sitter $\nu=\lambda$
we derive $y\left(  A_{1}\right)  =\frac{18\left(  1-\beta_{0}^{2}\right)
\left(  4\sqrt{3}\beta_{0}\lambda-3-4\lambda^{2}\right)  }{\left(  3-2\sqrt
{3}\beta\lambda\right)  ^{2}},~w_{eff}\left(  A_{1}\right)  =-1$ and
$e_{1}\left(  \lambda,\lambda,\beta_{0}\right)  =-6+\frac{3\left(
3-4\lambda^{2}\right)  }{3-2\sqrt{3}\beta_{0}\lambda}$. Therefore the de
Sitter point is an attractor when $\left\{  \beta_{0}\lambda<\frac{\sqrt{3}%
}{2},~\beta_{0}\lambda>\frac{4\lambda^{2}+3}{4\sqrt{3}},~\left\vert
\lambda\right\vert >\frac{\sqrt{3}}{2}\right\}  $ or$~\left\{  \beta
_{0}\lambda>\frac{\sqrt{3}}{2},~\beta_{0}\lambda<\frac{4\lambda^{2}+3}%
{4\sqrt{3}},~\left\vert \lambda\right\vert <\frac{\sqrt{3}}{2}\right\}  $, and
$\left\{  \left\vert \lambda\right\vert =\frac{\sqrt{3}}{2},~\beta_{0}%
\lambda\neq\frac{\sqrt{3}}{2}\right\}  $. \ For the second de Sitter solution
$\nu=\sqrt{3}\beta_{0}-\lambda$ we derive~$y\left(  A_{1}\right)  =-6,$
$w_{eff}\left(  A_{1}\right)  =-1$ and $e_{1}\left(  \lambda,\sqrt{3}\beta
_{0}-\lambda,\beta_{0}\right)  =-3$, from where we infer that the stationary
point is always an attractor and $y\left(  A_{1}\right)  <0$.

The stationary points $A_{2}$ and $A_{3}$ describe universes dominated by the
kinetic part of the scalar field, that is, $y\left(
A_{2}\right)  =0$ and $y\left(  A_{3}\right)  =0$. The points are real for
$\left\vert \beta_{0}\right\vert <1.$ The equation of state parameter is found
at each point $w_{eff}\left(  A_{2}\right)  =1-\frac{4}{3}\nu\left(  \beta
_{0}+\sqrt{\beta_{0}^{2}-1}\right)  \,$\ and $w_{eff}\left(  A_{2}\right)
=1-\frac{4}{3}\nu\left(  -\beta_{0}+\sqrt{\beta_{0}^{2}-1}\right)  $. In a
similar way as before for the linearized system near the stationary points we
find $e_{1}\left(  A_{2}\right)  =2\left(  3-\sqrt{3}\left(  \beta_{0}%
+\sqrt{\beta_{0}^{2}-1}\right)  \left(  \lambda+\nu\right)  \right)  $ and
$e_{1}\left(  A_{3}\right)  =2\left(  3+\sqrt{3}\left(  -\beta_{0}+\sqrt
{\beta_{0}^{2}-1}\right)  \left(  \lambda+\nu\right)  \right)  $. In the case
with $\nu=-\lambda$ we find that $e_{1}\left(  A_{2}\right)  =e_{2}\left(
A_{3}\right)  =6$, that is, the solutions are always unstable. Moreover, when
$\nu=\frac{\sqrt{3}}{2}\left(  \beta_{0}-\sqrt{\beta_{0}^{2}-1}\right)  $,
point $A_{2}$ describes a de Sitter universe which is stable when $\left\{
\lambda<-\sqrt{\frac{3}{2}},\frac{3+4\lambda^{2}}{4\sqrt{3}\lambda}<\beta
_{0}<-1\right\}  $, $\left\{  0<\lambda<\frac{\sqrt{3}}{2},~\beta_{0}%
<\frac{3+4\lambda^{2}}{4\sqrt{3}\lambda}\right\}  $ and $\left\{
\lambda>\frac{\sqrt{3}}{2},~\beta_{0}>1\right\}  $. Similarly, when $\nu
=\frac{\sqrt{3}}{2}\left(  \beta_{0}+\sqrt{\beta_{0}^{2}-1}\right)  $ point
$A_{3}$ describes a de Sitter solution which is stable when $\left\{
\lambda<\frac{\sqrt{3}}{2},\beta_{0}<-1\right\}  $, $\left\{  0<\lambda
<\frac{\sqrt{3}}{2},~\beta_{0}<\frac{3+4\lambda^{2}}{4\sqrt{3}\lambda
}\right\}  $ and $\left\{  \lambda>\frac{\sqrt{3}}{2},~1<\beta_{0}%
<\frac{3+4\lambda^{2}}{4\sqrt{3}\lambda}\right\}  $.

From these results it is clear that inflationary solutions are provided by
this cosmological model. From \eqref{st.23} it follows $V\left(  \psi\right)
=yA\left(  \psi\right)  H^{2}$. Thus, when $V\left(  \psi\right)  \neq0$ then
we shall say that the contribution of the aether field, through $A\left(
\psi\right)  ,$ is nonzero. That is true in the solution described by the
stationary point $A_{1}\,$, which can be seen as Lorentz violated inflationary
solution in the Jordan frame. Moreover, the asymptotic solutions described by
points $A_{1}$ and $A_{2}$ are similar with that of the usual scalar-tensor
theory \cite{ss}.

\subsection{Arbitrary potential}

For an arbitrary potential function, that is, for arbitrary function
$\Gamma\left(  \lambda\right)  $, then, the stationary points of the
two-dimensional dynamical system \eqref{st.25}-\eqref{st.26} are%
\begin{align*}
B_{1}\left(  \lambda_{0}\right)   &  =\left(  -\frac{3\beta_{0}-\sqrt{3}\nu
}{3-\sqrt{3}\beta_{0}\nu}, \lambda_{0}\right), \\B_{2}\left(  \lambda
_{0}\right) &  =\left(  -\beta_{0}+\sqrt{\beta_{0}^{2}-1},\lambda_{0}\right),\\
B_{3}\left(  \lambda_{0}\right)   &  =\left(  -\beta_{0}-\sqrt{\beta_{0}%
^{2}-1},\lambda_{0}\right)  \text{ and }\\
B_{4} & =\left(  0,\sqrt{3}\beta_{0}%
-\nu\right)  .
\end{align*}
in which $\lambda_{0}$ are solutions of the equation $\lambda^{2}\left(
\Gamma\left(  \lambda\right)  -1\right)=0$.

The family of stationary points $B_{1},$ $B_{2}$ and $B_{3}$ have the same
physical properties with that of points $A_{1},~A_{2}$ and $A_{3}$
respectively with $\lambda=0$. The new point $B_{4}$, is nothing else than the
de Sitter solution described by $A_{1}$ and $\nu=\sqrt{3}\beta_{0}-\lambda$.
What is different for an arbitrary potential is the stability of the points
which depends on the nature of function $\Gamma\left(  \lambda\right)  $. We
summarize the results in the following proposition.

\textit{Proposition 2: The stationary points for the Einstein-aether
scalar-tensor field equations \eqref{st.25}-\eqref{st.26} for arbitrary
potential, provide a de Sitter point as asymptotic solution, point }$B_{4}%
,$\textit{ and three families of scaling solutions described by points }%
$B_{1}$\textit{, }$B_{2}$\textit{ and }$B_{3}$\textit{ for every }$\lambda
_{0}$\textit{ which solves the equation }$\lambda_{0}^{2}\left(  \Gamma\left(
\lambda_{0}\right)  -1\right)  =0$\textit{. }

\subsubsection{Potential $V\left(  \psi\right)  =V_{0}\left(  e^{\sigma\psi
}-\Lambda\right)  $}

We demonstrate the results of the latter proposition by considering the
potential function $V\left(  \psi\right)  =V_{0}\left(  e^{\sigma\psi}%
-\Lambda\right)  $. For this potential we calculate $\Gamma\left(
\lambda\right)  =\frac{\sigma}{\psi}$, such that equation \eqref{st.26} to
become%
\begin{equation}
\frac{d\lambda}{d\ln a}=-2\sqrt{3}\lambda\left(  \lambda-\sigma\right)  x~.
\label{st.28}%
\end{equation}
Thus the polynomial equation $\lambda\left(  \lambda-\sigma\right)  =0$ has
the following roots, $\lambda_{1}=0$ and $\lambda_{2}=\sigma$. The stationary
points for the field equations \eqref{st.25}-\eqref{st.26} are%
\[
B_{1}\left(  0\right)  =\left(  -\frac{3\beta_{0}-\sqrt{3}\nu}{3-\sqrt{3}%
\beta_{0}\nu},0\right),\]  \[B_{2}\left(  0\right)  =\left(  -\beta_{0}%
+\sqrt{\beta_{0}^{2}-1},0\right),\]  \[B_{3}\left(  0\right)  =\left(
-\beta_{0}-\sqrt{\beta_{0}^{2}-1},0\right),
\]%
\[
B_{1}\left(  \sigma\right)  =\left(  -\frac{3\beta_{0}-\sqrt{3}\left(
\lambda+\nu\right)  }{3-\sqrt{3}\beta_{0}\left(  \lambda+\nu\right)  }%
,\sigma\right),\]  \[B_{2}\left(  \sigma\right)  =\left(  -\beta_{0}+\sqrt
{\beta_{0}^{2}-1},\sigma\right),\]  \[B_{3}\left(  \sigma\right)  =\left(
-\beta_{0}-\sqrt{\beta_{0}^{2}-1},\sigma\right)  .
\]
and%
\[
B_{4}=\left(  0,\sqrt{3}\beta_{0}-\nu\right)  .
\]

\begin{figure}[ptb]
\centering\includegraphics[width=0.8\textwidth]{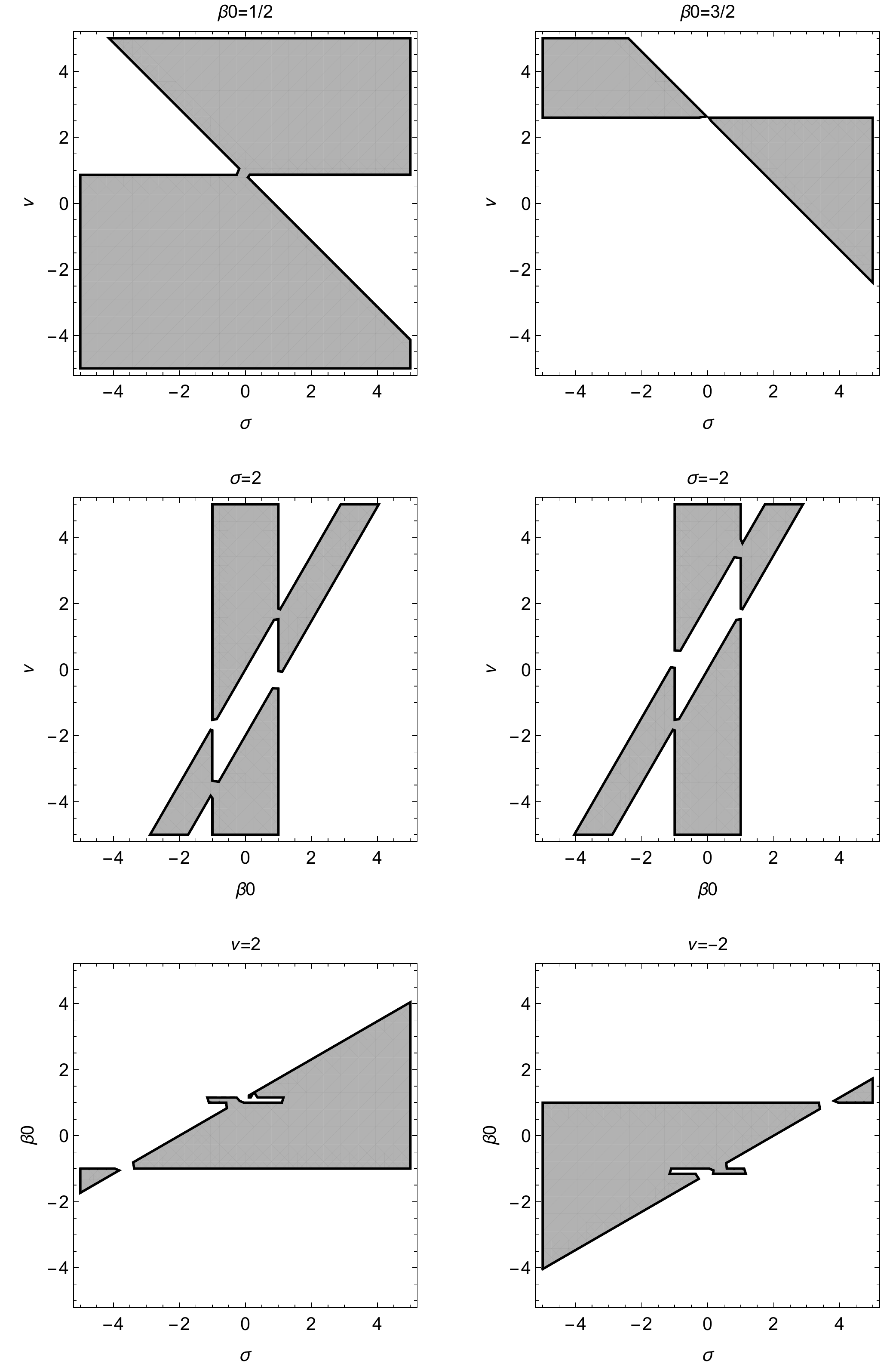}\caption{Region in
the spaces $\left\{  \sigma,\nu\right\}  $, $\left\{  \beta_{0},\nu\right\}  $
and $\left\{  \sigma,\beta_{0}\right\}  $ in which the real parts for the
eigenvalues $e^{\pm}\left(  B_{4}\right)  $ are negative and the de Sitter
universe described by $B_{4}$ is an attractor. }%
\label{fig1}%
\end{figure}

Let us focus on the stability conditions for point $B_{4}$. The eigenvalues of
the linearized system around the de Sitter point $B_{4}$ are%
\begin{equation}
e^{\pm}\left(  B_{4}\right)  =-\frac{3}{2}\pm\frac{\sqrt{3}}{2}\sqrt
{\frac{27\beta_{0}^{2}+8\nu\left(  \nu+\sigma\right)  -8\sqrt{3}\left(
2\nu+\sigma\right)  }{\sqrt{\beta_{0}^{2}-1}}.}%
\end{equation}
In Fig. \ref{fig1} we present region plots in the spaces $\left\{  \sigma
,\nu\right\}  $, $\left\{  \beta_{0},\nu\right\}  $ and $\left\{  \sigma
,\beta_{0}\right\}  $ in which the real parts for the eigenvalues
$e^{\pm}\left(  B_{4}\right)  $ are negative and the de Sitter universe
described by $B_{4}$ is an attractor.

\section{Symmetry classification}

\label{sec5}

In Section \ref{sec3} we determined a family of analytic solutions for
arbitrary functions $A\left(  \psi\right)  $ and $B\left(  \psi\right)  $ with
the empirical approach of reducing the field equations to a two-dimensional
system in Cartesians coordinates and recognizing its integrable form by our
experience in analytic mechanics. We now use a systematic method to
determine the scalar field potential $V\left(  \psi\right)  $ such that the
field equations to form a Liouville integrable system. The Noether symmetry
approach is applied in order to perform a classification of the scalar
field potential similar with that of the Ovsiannikov's classification
scheme.\ We briefly discuss the basic properties and definitions of Noether's theory.

Consider the one-parameter point transformation in the space of dependent
and independent variables $\left\{  t,x^{i}\left(  t\right)  \right\}  $:%
\begin{equation}
t^{\prime}=t+\varepsilon\xi\left(  t,x^{j}\right)  ~,~x^{i}=x^{i}%
+\varepsilon\eta^{i}\left(  t,x^{j}\right)   \label{ns.01}%
\end{equation}
and generator $Z=\xi\partial_{t}+\eta^{i}\partial_{i}$. Then, according to
Noether's first theorem, for a dynamical system provided by the variational
principle of the Lagrangian function $L\left(  t,x^{j},\dot{x}^{k}\right)  $,
the dynamical system remains invariant under the action of the
one-parameter point transformation with generator $X$ if and only if there
exists a function $f$ such that the following condition be true%
\begin{equation}
Z^{\left[  1\right]  }L+L\dot{\xi}=\dot{f}, \label{ns.02}%
\end{equation}
where $Z^{\left[  1\right]  }=Z+\left(  \dot{\eta}^{i}-\dot{x}^{i}\dot{\xi
}\right)  \partial_{\dot{x}^{i}}$ is the first extension of $X$ in the jet
space $\left\{  t,x^{i},\dot{x}^{i}\right\}  $. Condition \eqref{ns.02} is
called the Noether symmetry condition and, when it is true, the vector field
$Z$ is a Noether symmetry for the dynamical system described by the Lagrangian
function $L\left(  t,x^{j},\dot{x}^{k}\right)  $.

The importance for the derivation of Noether symmetries for a dynamical system
is two-fold. Noether symmetries form a subalgebra of the Lie point symmetries
of the dynamical system.  Thus the determination of Noether symmetries can be
used to define invariant functions or under the action of similarity
transformations to simplify the expression of the dynamical system. Moreover,
from Noether's second theorem there exists a formula which relates any Noether
symmetry to a specific conservation law.

Indeed, if $Z$ is the Noether symmetry for the dynamical system with
Lagrangian $L\left(  t,x^{j},\dot{x}^{k}\right),  $ then the following
quantity
\begin{equation}
\Phi=\xi\left(  \dot{x}^{k}\frac{\partial L}{\partial\dot{x}^{k}}-L\right)
-\eta^{k}\frac{\partial L}{\partial\dot{x}^{k}}+f \label{ns.03}%
\end{equation}
is a conservation law, i.e. $\frac{d\Phi}{dt}=0$.

The gravitational field equations of the Einstein-{\ae}ther scalar-tensor
cosmology form a two-dimensional Hamiltonian constraint system. The constraint
equation, which is the first modified Friedmann's equation can be seen as a
conservation law.  Thus we should determine a second conservation laws in order
to infer the integrability of the field equations for other forms of the
scalar field potential. We prefer to work in the coordinates $\left(
X,\Psi\right)  $ and the lapse function which we used to write the field
equations \eqref{st.17}-\eqref{st.19}. In these coordinates and for an arbitrary
potential function, the point-like Lagrangian is%
\begin{equation}
L\left(  X,\dot{X},\Psi,\dot{\Psi}\right)  =\left(  6\dot{X}^{2}+\frac{1}%
{2}\dot{\Psi}^{2}\right)  -\hat{V}\left(  \Psi\right)  e^{6X}, \label{ns.04}%
\end{equation}
where $d\Psi=\sqrt{1-3\beta\left(  \psi\right)  }d\psi$ and $\hat{V}\left(
\psi\right)  =V\left(  \psi\right)  A\left(  \psi\right)  e^{-3\int
\frac{B\left(  \psi\right)  }{A\left(  \psi\right)  }d\psi}$.

At this point it is important to mention that, while our original system was
dependent upon three free functions, $A\left(  \psi\right)  ,$~$B\left(
\psi\right)  $ and $V\left(  \psi\right),  $ we were able to find a specific
coordinate system in which only one function is essential, that is the scalar
field potential $V\left(  \psi\right)  $. Functions $A\left(  \psi\right)  $
and $B\left(  \psi\right)  $ play a significant role in the physical
properties of the solutions.  However, that is not true of the dynamics
described by \eqref{ns.04}. Function $A\left(  \psi\right)  $ has been
eliminated by the field equations by selecting a specific lapse function, while
$B\left(  \psi\right)  $ has been eliminated by a coordinate transformation on
the field $\psi$. Moreover, because we are interested on time-independent
conservation laws which are in-involution with the Hamiltonian, $\left\{
\Phi,\mathcal{H}\right\}  =0$, where $\left\{  \text{,}\right\}  $ is the Poisson
bracket and $\mathcal{H}=\dot{x}^{k}\frac{\partial L}{\partial\dot{x}^{k}}-L$,
from the results of the symmetry analysis of \cite{an21} we search for Noether
symmetries with $\xi\left(  t,x^{j}\right)  =0$.

We omit the derivation of the Lie point symmetries and the unique real
nonconstant scalar field potential $\hat{V}\left(  \Psi\right)  $ that we find
in which the point-like Lagrangian \eqref{ns.04} admits additional Noether
symmetries in the exponential potential $\hat{V}\left(  \Psi\right)  =\hat
{V}_{0}\exp\left(  -6\mu\Psi\right)  .$ The corresponding symmetry vector is
$Z=\partial_{X}+\frac{1}{\mu}\partial_{\Psi}$ while the resulting Noetherian
conservation law is $\Phi_{0}=12\dot{X}+\frac{1}{\mu}\dot{\Psi}$.

We define the new variable $X=\frac{u}{6}+\mu\Psi$.  Thus the point-like
Lagrangian \eqref{ns.04} is written as
\begin{equation}
L\left(  u,\dot{u},\Psi,\dot{\Psi}\right)  =\dot{u}^{2}+12\mu\dot{u}\dot{\Psi
}+3\left(  1+12\mu^{2}\right)  \dot{\Psi}^{2}-6\hat{V}_{0}e^{u},
\end{equation}
while the field equations are%
\begin{align}
\dot{u}^{2}+12\mu\dot{u}\dot{\Psi}+3\left(  1+12\mu^{2}\right)  \dot{\Psi}%
^{2}-6\hat{V}_{0}e^{u}  &  =0,\nonumber\\
\ddot{u}+3\hat{V}_{0}\left(  1+12\mu^{2}\right)  e^{u}  &  =0,\\
\ddot{\Psi}-6\hat{V}_{0}\mu e^{u}  &  =0.
\end{align}

Hence, the analytic solution is
\begin{equation}
u\left(  t\right)  =\ln\left(  \frac{u_{1}}{6\hat{V}_{0}\left(  1+12\mu
^{2}\right)  }\right)  -2\ln\left(  \cosh\left(  \frac{\sqrt{u_{1}}}%
{2}t\right)  \right)  ,
\end{equation}%
\begin{equation}
\Psi\left(  t\right)  =\frac{2}{3\left(  1+12\mu^{2}\right)  ^{2}}\left(
6\mu\left(  1+12\mu^{2}\right)  \ln\left(  \cosh\left(  \frac{\sqrt{u_{1}}}%
{2}t\right)  \right)  -\frac{\sqrt{3}}{2}\sqrt{\left(  -u_{1}\right)  }\left(
1+12\mu^{2}\right)  t\right)  \,.
\end{equation}

This is the analytic solution for the model the dynamics of which we studied in
Section \ref{sec4} when $\beta\left(  \psi\right)  =\beta_{0}$, $\Psi
\simeq\psi$ such that $\mu=\mu\left(  \lambda,\nu\right)  $ and function
$A\left(  \psi\right)  $ was the exponential function.

\subsection{Nonlocal symmetries}

The point-like Lagrangian \eqref{ns.04} is in a form similar to that of the
(phantom) minimally coupled scalar field $\Psi$ in general relativity where
$X$ plays the role of the scale factor, for the lapse function $N=X^{3}$. Thus,
if we select to work with nonlocal symmetries, we can obtain the results of
\cite{dm1} and end with the following proposition.

\textit{Proposition 3: The gravitational field equations of the
Einstein-{\ae}ther scalar-tensor with arbitrary potential functions }$A\left(
\psi\right)  $\textit{, }$B\left(  \psi\right)  $\textit{ and }$V\left(
\psi\right)  $\textit{ are integrable according to the existence of nonlocal
symmetries, consequently upon the existence of nonlocal conservation laws,
because of the existence of the constraint equation. }

The proof of Proposition 3 is straightforward, by applying the same
procedure with that of \cite{dm1}, thus we omit it. The result summarized in
the latter proposition is very important because if we elect to
solve the field equations numerically for any functional form of the three
unknown functions, we know that the numerical solutions correspond to
actual solutions.

\section{Conclusions}

\label{sec6}

In this work we introduced an Einstein-{\ae}ther scalar-tensor cosmological model
inspired by the Einstein-{\ae}ther scalar field model proposed in
\cite{Kanno:2006ty} in which the {\ae}ther coupling functions depend upon the scalar
field. In our proposed model we assumed that the scalar field is defined in
the Jordan frame and that it is coupled to gravity. In the case of a spatially flat
FLRW background space the field equations describe the evolution of the scale
factor and of the scalar field, while there are three unknown functions which
should be defined. These functions are related with the scalar field
potential, the scalar-tensor coupled function to gravity and the {\ae}ther
coefficient functions. The field equations are described by a minisuperspace
and a point-like Lagrangian.  This is an important observation because
important techniques from Analytic Mechanics can be applied.

We investigated the integrability property of the field equations and the
existence of analytic solutions. We were able to define a new set of variables
in the minisuperspace and to eliminate two of the three unknown functions of
the model. We end with a dynamical system in which only the scalar field
potential drives the dynamics. The other two unknown functions have been
eliminated by a change in the lapse function of the background space and by a
coordinate transformation. These functions play a significant role in the
physical properties of the cosmological solution, but they do not affect the
dynamics in the new variables.

In order to understand the evolution of the physical variables, we performed a
detailed study of the field equations by investigating the stationary points in
the dimensionless variables of the $H$-normalization approach. For exponential
functional forms of the coupled variables and exponential scalar field
potential, we found that there exists three stationary points which in general
describe scaling solutions.\ The effective cosmological fluid in two of the
stationary points depends only upon the kinetic part of the scalar field, while
the scalar field potential contributes in the third point. These solutions can
be seen as analogues in the Jordan frame of the Lorentz-violation inflationary
solutions which were found in \cite{Kanno:2006ty} defined in the Einstein
frame. Moreover, for an arbitrary potential function, we found that the
stationary points belong to the three families described by the exponential
potential with an addition a new stationary point which provides a de Sitter universe.

Finally, we performed a symmetry classification for the field equations
according to the admitted Noether point symmetries, inspired by
Ovsiannikov's classification scheme. We found that for the scalar field
potential, the field equations admit additional conservation laws and the
cosmological solution can be written in closed-form expression. In addition,
this closed-form solution can describe the generic evolution which was found
by the analysis of the stationary points in the $H$-normalization. That is a
very interesting result.  We can relate the analysis of the stationary points
with an actual solution for the dynamical system.

From this analysis on the extension of the Einstein-aether scalar field model
in the Jordan frame, we can conclude that the cosmological model can be
physically viable since it can provide important eras in the cosmological
evolution. In a future study we plan to investigate further the physical
properties of the model as also to investigate the effects of the conformal
transformation in the gravitational Action Integral.

\begin{acknowledgments}
This work is based on the research supported in part by the National Research
Foundation of South Africa (Grant Number 131604). AP \& GL were funded by
Agencia Nacional de Investigaci\'{o}n y Desarrollo - ANID through the program
FONDECYT Iniciaci\'{o}n grant no. 11180126. Additionally, GL is supported by
Vicerrector\'{\i}a de Investigaci\'{o}n y Desarrollo Tecnol\'{o}gico at
Universidad Cat\'olica del Norte.
\end{acknowledgments}

\end{document}